\documentclass{ws-procs975x65}

\begin{document}

\title{$\Delta-$String: A Hybrid Between Einstein's and String Paradigmes}

\author{Vladimir Dzhunushaliev\footnote{\uppercase{W}ork 
supported by \uppercase{A}lexander von  
\uppercase{H}umboldt \uppercase{F}oundation.}}

\address{Dept. Phys. and Microel. Engineer., KRSU, Bishkek, \\
Kievskaya Str. 44, 720021, Kyrgyz Republic\\ 
E-mail: dzhun@hotmail.kg}

\author{Hans-J\"urgen Schmidt\footnote{\uppercase{V}aluable comments 
by \uppercase{E. M}ielke are gratefully acknowledged. 
\uppercase{T}hanks the \uppercase{D}eutsche \uppercase{F}orschungsgemeinschaft 
\uppercase{DFG} for financial support.}}

\address{Universit\"at Potsdam, Institut f\"ur Mathematik,
14469, Potsdam, Germany \\
E-mail: hjschmi@rz.uni-potsdam.de}  


\maketitle

\begin{figure}[ht]
 \epsfxsize=4cm
 \begin{center}
   \fbox{
   \epsfbox{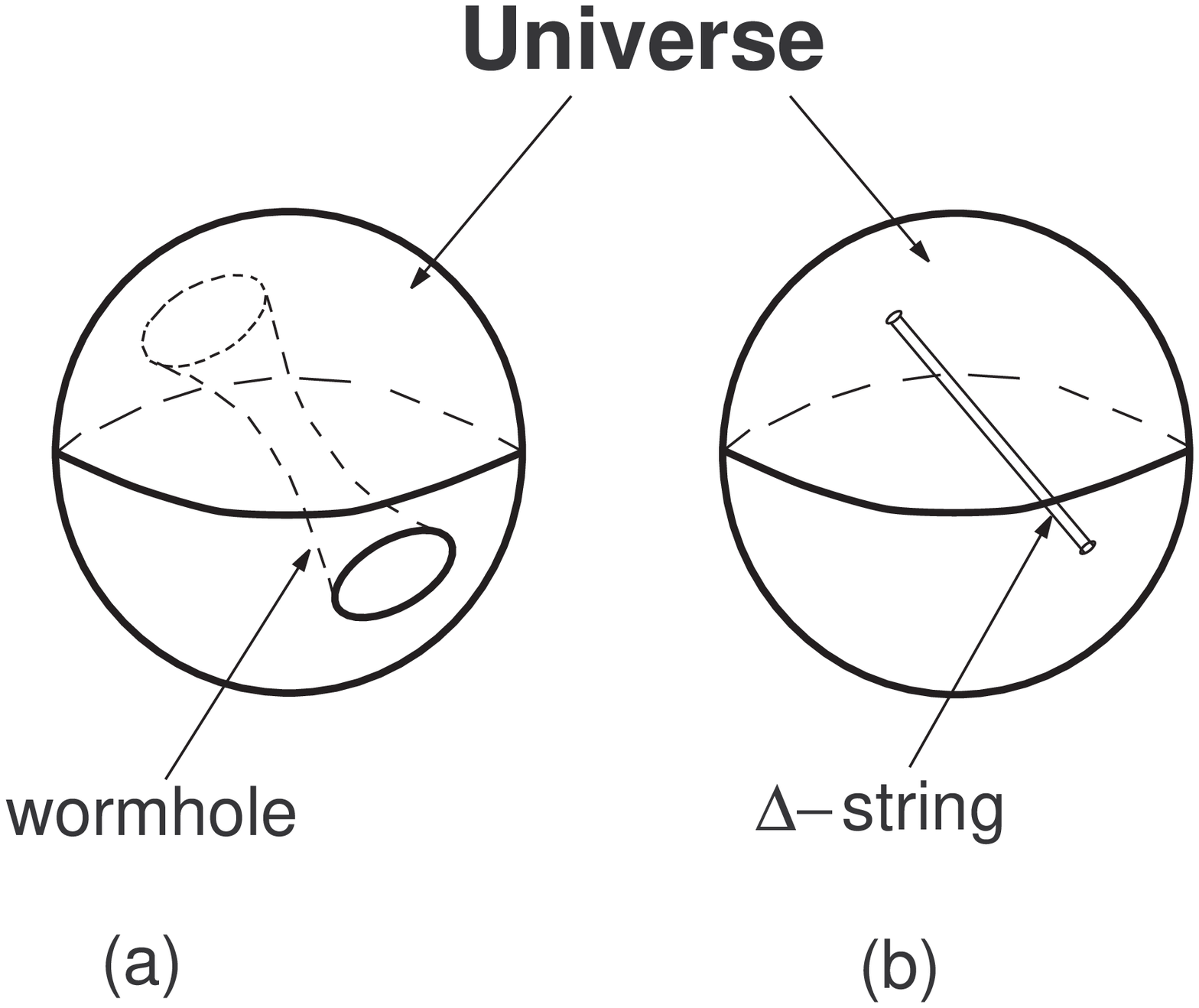}}
 \end{center}
 \caption{The wormhole \textbf{(a)} or thin gravitational flux tube 
 \textbf{(b)} attached to remote parts of a single Universe.}
 \label{fig1}
\end{figure}

A. Einstein had an idea that the electron should have an inner
 structure which allows us to avoid singularities of point-like 
particles in classical and quantum electrodynamics. 
Einstein's-Wheeler's idea about this structure is that the electron 
is a wormhole: a bridge connecting two remote parts of a Universe. 
The picture for this point of view is presented on Fig.\ref{fig1}a. 
\par 
One of the bad peculiarities of this point of view is that the wormhole 
has a curvature in the same order as an entire part of the Universe. 
It means that the linear sizes of the wormhole can be compared 
with the sizes of the Universe and the whole object $=$
 Universe $+$ wormhole is too curved. But may be, the picture 
presented on Fig.\ref{fig1}b is more attractive. 
Here the cross section of the wormhole is of the order of  the Planck
length. In this case, the external part of the Universe is as it is, 
and the attachment points are like to $(+)$ and $(-)$ electric charges. 
We see that such a bridge is more similar to a string attached to a 
D-brane. 
\par
Thus, our aim is to show that in vacuum 5D gravity there
 exist wormhole-like solutions with a superthin and superlong mouth. 
We call such a wormhole  a $\Delta$-string with the following 
motivation for the words $\Delta$ and string:
(i)``string" means that it is very thin and very long,
(ii) $\Delta$ means that the attachment point is like  a delta
     of a river in the consequence of a spacetime foam.
In other words, the attachment point of the $\Delta$-string is spread. 
\par
The initial equations for the description of the $\Delta$-string are 
ordinary 5D Einstein's equations. 
The detailed investigation of all spherically symmetric solutions tells
 us that the solutions depend on the relation between electric 
$q$ and magnetic $Q$ charges, and the case most interesting for us is 
$(1 - Q/q) << 1$ where $Q<q$. 
The numerical investigation shows us that in this case the solution is 
similar to Fig.\ref{fig1}b: there is an arbitrarily long mouth with the  
length $l \rightarrow \infty$ by $\delta \rightarrow 0$. 
\par
The numerical and approximate calculations \cite{vd1} \cite{vd2} show that:
(i) $a(r_H) = 2 a(0)$, $r_H$ is defined from the relation: 
$ds^2(\pm r_H) = 0$; 
(ii) $l(\delta ) \approx a(0) \ln \delta $.
The cross section of the $\Delta$-string of the origin $a(0)$ is arbitrary, 
and can be chosen  $a(0) \approx l_{Pl}$. 
Thus the mouth of the 5D spherically symmetric wormhole-like solution with 
$q \approx Q$ \ $(q > Q)$ can be superthin and superlong, and 
consequently can be considered as a string-like object, namely 
a $\Delta$-string.
\par 
It is necessary to note that one can insert this part of a flux tube 
solution between two Reissner-Nordstr\"om black holes. This situation is much 
more interesting because it is like to a string attached to 2 D-branes : 
the flux tube solution is the string, each Reissner-Nordstr\"om solution 
is D-brane and joining on the event horizon takes place.
\par 
Next,  we would like to reduce our initial 5D Lagrangian to 
a  2D Lagrangian. 
At this step we set the sizes of 5$^{\mathrm{th}}$ and $S^2$ dimensions 
$\approx l_{Pl}$.  The first step is the usual 5D $\rightarrow$ 4D 
Kaluza-Klein dimensional reduction. 
The further step is the  reduction from 4D to 2D. We consider 
the region of spacetime where the topology is 
$M^2 \times S^2 \times S^1$ and the linear sizes of $S^2$ 
($\theta, \varphi coordinates$) are 
$\approx l_{Pl}$. We can give arguments that in this case all 
physical fields do not depend on the coordinates on 
the sphere $S^2$. The 4D metric can be expressed as 
\begin{eqnarray}
d\stackrel{(4)}{s^2} & = & g_{\mu \nu} dx^\mu dx^\nu = 
g_{ab}(x^c) dx^a dx^b + 
\nonumber \\
&&\chi(x^c)
\left (
\omega ^{\bar i} + B^{\bar i}_a(x^c) dx^a
\right )
\left (
\omega _{\bar i} + B_{\bar i a}(x^c) dx^a
\right )
\label{sec2-30}
\end{eqnarray}
where $a,b = 0,1$; $x^a$ are the time and longitudinal coordinates; 
$-\omega^{\bar i} \omega_{\bar i} = dl^2$ is the metric 
on the 2D sphere $S^2$. The dimensional reduction to 2 dimensions is 
\begin{eqnarray}
\stackrel{(4)}{R} & = & \stackrel{(2)}{R} + R(S^2) - 
\frac{1}{4} \Phi^{\bar i}_{ab} \Phi^{ab}_{\bar i} - 
\nonumber \\
&&\frac{1}{2} h^{ij}h^{kl}
\left (
D_a h_{ik} D^a h_{jl} + D_a h_{ij} D^a h^{kl}
\right ) - 
\nabla^a 
\left (
h^{ij} D_a h_{ij}
\right )
\label{sec2-40}
\end{eqnarray}
where $\stackrel{(2)}{R}$ is the Ricci scalar of 2D spacetime; 
$D_\mu$ and $\Phi^{\bar i}_{ab}$ are, respectively, the covariant derivative 
and the curvature of the principal connection $B^{\bar i}_a$, and 
$R(S^2)$ is the Ricci scalar of the sphere $S^2$; $h_{ij}$ is some metric.
\par 
Let us consider the situation with the electromagnetic 
fields $A_\mu$ and $F_{\mu \nu}$. 
\begin{eqnarray}
A_\mu & = & \left \{ A_a, A_i \right \} \; 
A_a \; \text{is the vector;} \; 
A_i \; \text{are 2 scalars}; 
\label{sec2-80}\\
F_{ab} & = & \partial_a A_b - \partial_b A_a \; 
\text{is the Maxwell tensor for} \; A_a;
\label{sec2-90}\\
F_{ai} & = & \partial_a A_i ,
\label{sec2-100}\\
F_{ij} & = & 0
\end{eqnarray}
here we took into account that $\partial_i = 0$. 
\par 
Connecting all results we see that only the following 
\textit{physical fields} on the ($\Delta$-string) are possible : 
2D metric $g_{ab}$, gauge fields $B^{\bar i}_a$, vectors $A_a$,
tensors $F_{ab}, F_{a\bar i}, \Phi^{\bar i}_{ab}$, scalars $\chi$ and 
$\phi$. 
\par
The next remark is that $\Delta$-string can have an essential decreasing 
of the initial degrees of freedom. For example, $G_{\theta \theta}$ 
and $G_{\phi \phi}$ describe the lengths in the transversal 
directions but these directions are in the Planck region and 
consequently these degrees of freedom should be excluded from 
the effective Lagrangian: 
$\sqrt{-G} R_{(5)} \rightarrow L_{eff}$
with the loss of some degrees of freedom. 
\par 
In conclusion, the $\Delta$-string has the next very interesting properties 
\begin{enumerate}
     \item In during of one of the 
paradigm of quantum gravity that the spacetime has not any 
structure by the length less Planck length we see that some 
vacuum solutions in the Kaluza-Klein gravity actually 
do not have any transversal structure, i.e.
the $\Delta$-string is really a 1-dimensional object.
     \item The $\Delta$-string has a flux of electric field and 
consequently   for an external observer living on the D-brane 
it looks like to a spread of  electric charge.
     \item The $\Delta$-string has the mixed
classical and quantum properties: in the longitudinal direction 
it has classical properties but in transversal a quantum one. 
     \item
     The $\Delta$-string is an object where we can 
     investigate the 2D spacetime foam.  
\end{enumerate}
The next investigations in this direction will be based on the 7D metrics 
presented in \cite{hjss}-\cite{ds1}.

\end{document}